\documentclass[prl,onecolumn,preprint,superscriptaddress,showpacs]{article}
\usepackage{epsfig,cite}
\usepackage{amsmath,amssymb,graphicx}
\usepackage{ae}
\usepackage{aecompl}
\usepackage[margin=3.0cm,bottom=3.5cm]{geometry}

\begin{document}

\title{Kinetics of copolymer localization at a selective liquid-liquid interface}
\author{A. Corsi$^{1}$, A. Milchev$^{1,2}$, V.G. Rostiashvili$^1$, and 
T.A. Vilgis$^1$\\
$^1$Max - Planck - Institute for Polymer Research -
Ackermannweg 10,\\ 55128 Mainz, Germany\\
$^2$Institute for Physical Chemistry, Bulgarian Academy of
  Sciences,\\ 1113 Sofia, Bulgaria
}
%\pacs{36.20.-r} \pacs{68.10.-m} \pacs{07.05.Tp}

\maketitle

\begin{abstract}
The localization kinetics of a regular block-copolymer of total length $N$ and block size $M$ at a selective liquid-liquid interface is studied in the limit of strong segregation between hydrophobic and polar segments in the chain.
We propose a simple analytic theory based on scaling arguments which describes the relaxation of the initial coil into a flat-shaped layer for the cases of both Rouse and Zimm dynamics. For Rouse dynamics the characteristic times for attaining equilibrium values of the gyration radius components perpendicular and parallel to the interface are predicted to scale with block length $M$ and chain length $N$ as $\tau_{\perp} \propto M^{1+2\nu}$ (here $\nu\approx 0.6$ is the Flory exponent) and as $\tau_{\parallel} \propto N^2$, although initially the characteristic coil flattening time is predicted to scale with block size as $\propto M$. Since typically $N\gg M$ for multiblock copolymers, our results suggest that the flattening dynamics proceeds faster perpendicular rather than parallel to the interface, in contrast to the case of Zimm dynamics where the two components relax with comparable rate, and proceed considerably slower than in the Rouse case.

We also demonstrate that, in the case of Rouse dynamics, these scaling predictions agree well with the results of Monte Carlo simulations of the localization dynamics. A comparison to the localization dynamics of {\em random} copolymers is also carried out. 
\end{abstract}

\section{Introduction}
The behavior of hydrophobic - polar (HP) copolymers at a selective penetrable interface (the interface which divides two immiscible liquids, like water and oil, each of them being favored by one of the two types of monomers) is of great importance in the chemical physics of polymers. For strongly selective interfaces, when  the energy gain for a monomer in the favored solvent is large, the hydrophobic and polar blocks of a copolymer chain try to stay on different sides of the interface leading thus to a major reduction of the interfacial tension between the immiscible liquids or melts which has important technological applications, e.g. for compatilizers, thickeners or emulsifiers. Not surprisingly, during the last two decades the problem has gained a lot of attention from experiment \cite{Clifton,Rother,Wang,Omarjee}, theory \cite{Sommer1,Sommer2,Garel,Joanny,Denesyuk} as well as from computer experiment \cite{Balasz,Israels,Sommer3,Lyats,Chen}. While in earlier studies attention has been mostly focused on diblock copolymers \cite{Rother,Omarjee} due to their relatively simple structure, the scientific interest shifted later to {\em random} HP-copolymers at penetrable interfaces \cite{Sommer2, Garel,Joanny,Denesyuk,Chen}. Until recently though, the properties of regular multiblock copolymers, especially with emphasis on their {\em dependence on block length} $M$, have remained largely unexplored. This has been a driving force for our research, since experimentalists are now able to synthesize polymers with an increased variety of controlled structures.

In a recent study \cite{Corsi}, we showed that the equilibrium properties (structure, diffusion coefficient, etc.) of a regular HP-copolymer at a selective liquid-liquid interface are well described by a scaling theory in terms of  the total copolymer length $N$ (the number of repeating units in the chain) and the block size $M$ (the number of consecutive monomers of the same kind) as well as the selectivity parameter $\chi$, that is, the energy gained by a monomer when being in the more favorable solvent. 

In terms of the selectivity parameter, there are three adsorption regimes that can be distinguished: 
\begin{itemize}
\item if $\chi$ is smaller than a critical value $\chi_{c}$, the interface is too {\it weak} to affect the 
polymer at all so that the macromolecule conformation is identical to that in the absence of an interface;
\item for $\chi > \chi_{c}$ the polymer starts to feel the presence of the interface so that  the coil 
radius perpendicular to the interface starts to shrink while parallel to the interface this radius swells. 
In this regime the scaling approach developed in \cite{Corsi} correctly describes the statics of the 
problem. This regime is labeled as of {\em weak localization};
\item if $\chi$ is increased furthermore, a second threshold value $\chi_{\infty}$ is reached. This 
value corresponds to an interface that is strong enough to induce a {\em perfect flattening} of the 
copolymer in which all the monomers are in their preferred environment. In this regime of {\em 
strong localization}, an increase in $\chi$ does not change the shape of the polymer anymore since 
the polymer is already in equilibrium, as flat as it can get, depending on its structure, mainly 
on the length of the blocks.
\end{itemize}

Using scaling arguments, we demonstrated \cite{Corsi} that: (i) the crossover selectivity decreases 
with growing block length as $\chi_c \propto M^{-(1+\nu)/2}$ (where $\nu\approx 0.6$ is the Flory 
exponent in three dimensions) and the threshold to the strong localization regime vanishes as $\chi_{\infty} 
\propto M^{-1}$, and (ii) the size of the copolymer varies in the weak localization regime as $R_{g\perp} 
\propto M^{-\nu(1+\nu)/(1-\nu)}$ and $R_{g\parallel} \propto M^{(\nu_2 - \nu) (1+\nu)/(1-\nu)}$, whereas 
$R_{g\perp}\propto M^\nu,\quad R_{g\parallel} \propto M^{-(\nu_2-\nu)}$ in the case of strong localization 
(where $\nu_2 = 3/4$ is the Flory exponent in two dimensions). We shall use these results in the present 
paper in which we suggest a theory of the localization kinetics of a regular block copolymer at a 
liquid-liquid interface based on dynamical scaling arguments. To the best of our knowledge, so far 
there have been no attempts to treat  this problem analytically or by means of computer simulation.

\section{Scaling analysis}

In our {\em dynamical scaling analysis} we consider a coarse-grained model of a multiblock copolymer consisting of $N$ repeat units which is built up from a sequence of H- and P - blocks each of length $M$. For simplicity one may take the interface with negligible thickness as a flat plane which separates the two selective immiscible solvents.  The energy gain of each repeat unit is thus $-\chi$, provided it stays in its preferred solvent, and the system is considered in the {\em strong localization} limit where $\chi >\chi_{\infty} \propto M^{-1}$ \cite{Corsi,Leclerc}. In our consideration, we skip the time needed by the polymer to move by diffusion through the bulk into the vicinity of the interface, and place the center of mass of an unperturbed coil at time  $t = 0$ at the interface whereby the localization field is switched on.  The initial coil will then start relaxing with time into a flat (``pancake'') equilibrium configuration in-plane with the interface and the kinetics of relaxation will be determined by the sum of the various forces acting on the copolymer.

It might be argued that a more {\em physical} initial configuration could be represented by a 
polymer that, after diffusing in the bulk of one of the solvents, gets to the interface and 
touches it with one single monomer. We have indeed tried this initial configuration as well,
and concluded that, once the adsorption process starts, it proceeds exhibiting the same physics 
(e.g. the same scaling behavior, see Section 4.1). The choice of our initial condition just 
eliminates the spread in the initial time at which adsorption sets on. This spread is caused
by polymers which occasionally drift away from the interface, provided the coil touches the
interface initially with the ``wrong'' kind of P- or H-monomers.
 
\subsection{Rouse dynamics}
In order to estimate the driving force of the flattening process, one may recall \cite{Corsi,Leclerc} that the effective attractive energy $\epsilon$ (per diblock, e.g. a pair of consecutive blocks of H- and P- monomers) in the strong localization limit is $ \epsilon \approx \chi M $. In this case the diblock plays the role of a blob and the overall attractive free energy $ F_{\rm attr} \propto \epsilon {\cal N} \propto \chi N$ where ${\cal N}\simeq N/M $ is the total number of blobs. Thus  the effective driving force perpendicular to the interface is $f_{\rm attr}^{\perp} \approx -\chi_{\infty}N/R_{\perp}$ where $R_{\perp}$ denotes the perpendicular component of the radius of gyration. This force is opposed by a force of confinement due to the deformation of the self-avoiding chain into a layer of thickness $ R_{\perp} $. The corresponding free energy of deformation is simply estimated as $ F_{\rm conf} \simeq N(b/R_{\perp})^{1/\nu}$ where $b$ is the Kuhn segment size \cite{Gennes}. For the respective force then one gets $f_{\rm conf}^{\perp} \simeq Nb^{1/\nu}/ R_{\perp}^{1/\nu + 1}$.

The equation of motion for $R_{\perp}$ follows from the condition that the friction force which the chain experiences during the motion in the direction perpendicular to the interface is balanced by the sum of $f_{\rm attr}^{\perp}$ and  $f_{\rm conf}^{\perp}$. In the case of Rouse dynamics, each chain segment experiences independent Stokes friction so that the resulting equation of motion has the form
\begin{eqnarray}
\zeta_{0} N \frac{d R_{\perp}}{d t}  = - \frac{\chi_{\infty}N}{R_{\perp}} + \frac{N b^{1/\nu}}{R_{\perp}^{1/\nu + 1}} \quad,
\label{Eq_motion_perp}
\end{eqnarray}
where $\zeta_{0}$ is the friction coefficient per segment.

During the flattening process, the chain spreads parallel to the interface due to the excluded volume interaction. Within Flory mean - field arguments, the corresponding free energy $F_{\rm ev} \simeq v N^2 / (R_{\parallel}^2 R_{\perp})$, where $v$ is the second virial coefficient and $R_{\parallel}$ is the gyration radius component parallel to the interface.
The corresponding driving force is $f_{\rm ev}^{\parallel} \simeq v N^2 / (R_{\parallel}^3 R_{\perp})$. This term is counterbalanced by the elastic force of chain deformation, i.e. by $f_{\rm def}^{\parallel} \simeq  - R_{\parallel}/(b^{2} N)$. 

Taking into account the balance of forces in the parallel direction, the equation of motion for $R_{\parallel}$ takes then the following form
\begin{eqnarray}
\zeta_{0} N \frac{d R_{\parallel}}{d t}  = \frac{v N^2}{R_{\parallel}^3 R_{\perp}}  - \frac{R_{\parallel}}{b^{2} N} \quad.
\label{Eq_motion_parallel}
\end{eqnarray}
Evidently, the excluded volume interactions provide a coupling between the relaxation perpendicular and parallel to the interface.
Thus, eqs. (\ref{Eq_motion_perp}) and (\ref{Eq_motion_parallel}) describe the relaxation kinetics of a multiblock copolymer conformation at a selective liquid-liquid interface.
One may readily verify that the equilibrium solutions which follow from these equations are
\begin{eqnarray}
\label{Equilibrium_perp}
R_{\perp}^{\rm eq} \simeq b M^{\nu},
\end{eqnarray}
and
\begin{eqnarray}
\label{Equilibrium_paral}
R_{\parallel}^{\rm eq} \simeq (v b)^{1/4} M^{-\nu/4} N^{\nu_{2}}.
\end{eqnarray}
These coincide with the equilibrium expressions for $R_{\perp}$ and $R_{\parallel}$ derived earlier from purely scaling-based considerations \cite{Corsi}. The only difference is in the power of the $M$ - dependence in eq. (\ref{Equilibrium_paral}) which looks like $M^{-\nu/4}$ instead of $M^{-(\nu_{2} - \nu)}$ in ref. \cite{Corsi}, albeit numerically the values of both exponents coincide: $\nu/4 \simeq (\nu_{2} - \nu) \simeq  0.15$. To get the full solution of the equations of motion it is convenient to rescale the variables as $x \equiv R_{\perp}/(b M^{\nu}),\quad y \equiv R_{\parallel}^{4} M^{\nu}/(v b N^3)$ so that eqs. (\ref{Eq_motion_perp}) and (\ref{Eq_motion_parallel}) can be written in the dimensionless form
\begin{eqnarray}
\tau_{\perp} \: \frac{d x}{d t} &=& \frac{1}{x^{1/\nu + 1}} - \frac{1}{x} \label{Eq_motion_x}\\
\tau_{\parallel} \: \frac{d y}{d t} &=& \frac{1}{x} - y \quad.
\label{Eq_motion_y}
\end{eqnarray}

The characteristic times for relaxation perpendicular and parallel to the interface in eqs. (\ref{Eq_motion_x}) - (\ref{Eq_motion_y}) should then scale as $ \tau_{\perp} \simeq \zeta_0 b^2 M^{1 + 2\nu} $, and $\tau_{\parallel} \simeq \zeta_0 b^2 N^2 $.

Eq. (\ref{Eq_motion_x}) is independent of $y$ so that it can be solved exactly:
\begin{eqnarray}
\frac{x^2(t)}{2}\left[ 1 - F(2 \nu,1;1 + 2 \nu; x^{1/\nu}(t))\right] \nonumber \\
- \frac{x^2(0)}{2}
\left[ 1 - F(2 \nu,1;1 + 2 \nu; x^{1/\nu}(0))\right] = - \frac{t}{\tau_{\perp}} \quad,
\label{Solution}
\end{eqnarray}
where $ F(\alpha, \beta; \gamma; z) $ is the hypergeometric function and $ x(0) = R_{\perp}(0)/b M^{\nu} $ is the initial value. In the {\em early stages} of relaxation (i.e. at $t \ll \tau_{\perp}$) one has $x \gg 1$ and the solution, eq. (\ref{Solution}), with $ R_{\perp}(0) \simeq b N^{\nu} $ reduces to
\begin{eqnarray}
R_{\perp}^{2}(t)  \simeq  R_{\perp}^{2}(0) - \frac{t}{\zeta_{0} M} \quad,
\label{Solution_limit11}
\end{eqnarray}
so that the perpendicular collapse of the chain should last proportionally to the block length $M$.

In the opposite limit of {\em late stages} kinetics, $t \simeq \tau_{\perp}$ and $x \geq 1$, that is, close to equilibrium, the relaxation of $R_{g}(t)$ is essentially exponential with $\tau_{\perp}\propto M^{2.2}$:
\begin{eqnarray}
\frac{R_{\perp}(t)}{ R_{\perp}^{\rm eq} } \simeq 1 +\exp(-\frac{t}{\tau_{\perp}})\quad.
\label{Solution_limit2}
\end{eqnarray}

\begin{figure}[ht] 
\begin{center}
\epsfig{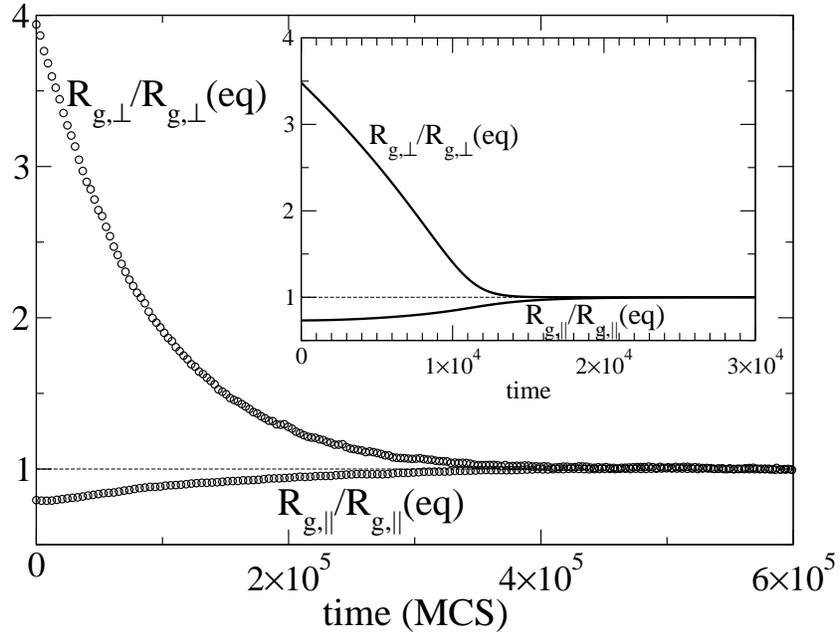}
\vskip 1.0cm
\caption{\label{analyt_vs_MC}
Relaxation of $R_\perp$ and $R_\parallel$ with time, following from Monte Carlo simulations 
 of a copolymer of length $N = 128$ and
block size $M = 16$. Both components are normalized by their respective equilibrium
values. The inset shows the result for the same quantities (and the same set of
parameters) as obtained from the numerical solution of eqs. (\ref{Eq_motion_x}) and (\ref{Eq_motion_y}).}
\end{center}
\end{figure}

Moreover, as typically $ N \gg M $, one expects that $\tau_{\parallel} \gg \tau_{\perp}$, i.e. the chain coil collapses first in the perpendicular direction to its equilibrium value and after that slowly extends in the parallel direction. In this case one can use $ x(t) \simeq x^{\rm eq} =1 $ in eq.(\ref{Eq_motion_y}) and derive the resulting solution for the parallel component of $R_g$ as
\begin{eqnarray}
y(t) - 1 = \left[y(0) - 1\right]\exp(-t/\tau_{\parallel}) \quad,
\label{Solution_y}
\end{eqnarray}
where the initial value $ y(0) < 1 $.
As one may see in Fig. \ref{analyt_vs_MC}, the time evolution of $R_\perp$ and $R_\parallel$ which follows 
from the solution of the system of differential equations (\ref{Eq_motion_x}) - (\ref{Eq_motion_y}), resembles 
qualitatively the simulation data even though one should bear in mind that the time axis for the former is 
in arbitrary units (scaling always holds up to a prefactor) whereas in the simulation time is measured in 
MC steps (MCS) per monomer (i.e., a MCS has passed after all monomers have been allowed to perform a move 
at random). The relaxation of $R_\parallel$ looks somewhat slower and, as our data for longer chains show, 
this effect becomes much more pronounced with growing number of blocks.

Concerning Fig. \ref{analyt_vs_MC}, note that there are several approximations on which our analytical derivations
rely. First of all, our expressions for the attractive free energy,  $F_{\rm attr}$, and the confinement free 
energy, $F_{\rm conf}$, are only appropriate in the limit of strong flattening which occurs during the late 
stages of the localization process.  Moreover, the free energy $F_{\rm ev}$, due to excluded volume 
interactions, is somewhat underestimated if one uses the second virial term approximation only.  That 
explains why both analytic curves in Fig. \ref{analyt_vs_MC}  qualitatively agree well with the simulation data 
at the late stages of the relaxation process whereas, not surprisingly, this agreement is missing during 
its initial stages.

\subsection{Zimm dynamics}

\begin{figure}[ht]
\begin{center}
\epsfig{file=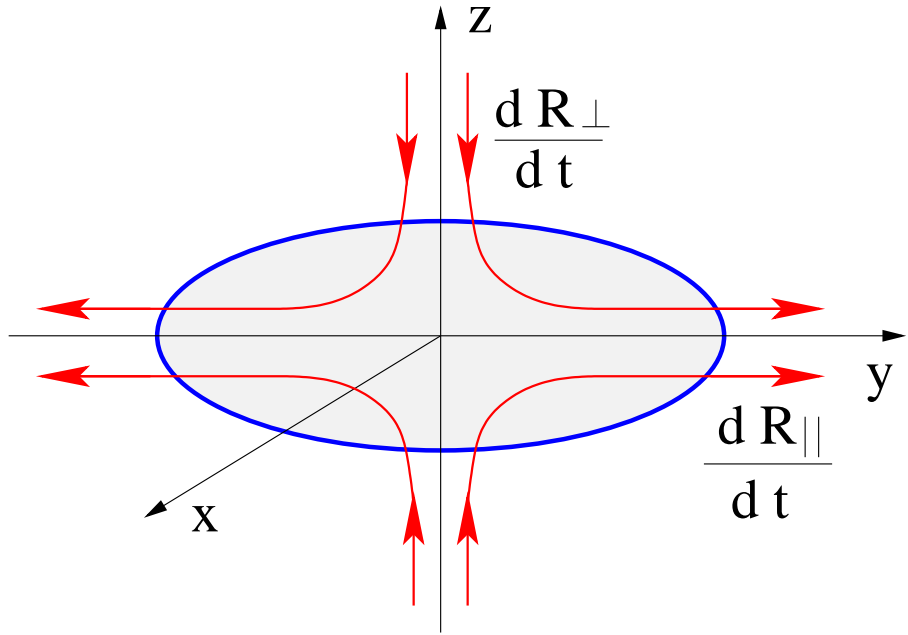,width=12cm}
\caption{The solvent velocity profile in the Zimm model due to the entrainment of the 
solvent by the coil segments. The corresponding velocity gradients are $\nabla_z  
u_x =  \nabla_z  u_y \simeq (d R_{\parallel}/d t)/R_{\perp}$ and $ \nabla_x u_z =  
\nabla_y  u_z \simeq (d R_{\perp}/d t)/R_{\parallel}$.}
\label{Zimmpicture}
\end{center}
\end{figure}

In the case of Zimm dynamics, the hydrodynamic long-ranged interactions couple the motion of the solvent to that of the coil within the framework of the non-draining coil model \cite{Grossberg}. When the coil  flattens, the solvent velocity $ {\bf u}(r) $ follows the velocity of the coil segments (the solvent entrainment), as depicted schematically in Fig. \ref{Zimmpicture}. In order to calculate the friction forces 
which act during the coil flattening, we recall that the energy dissipation in an incompressible fluid is \cite{Landau}:
\begin{eqnarray}
\dot{E} \simeq - \eta_0 \int d^3r \left (\nabla_k  u_i(r) + \nabla_i u_k(r)\right )^2 \quad,
\label{Dissipation}
\end{eqnarray}
where $ {\bf u}(r) $ is the solvent velocity field, $ \eta_0 \sim \zeta_0/b $ is the viscosity, and the integral is taken over the interior of the coil. In our case the energy dissipation can be estimated as \begin{eqnarray}
\dot{E} &\simeq& - \eta_0 \left(\frac{1}{R_{\parallel}}\frac{d R_{\perp}}{d t}  +  
\frac{1}{R_{\perp}}\frac{d R_{\parallel}}{d t}\right)^2 R_{\parallel}^2 
R_{\perp}\nonumber\\
&\simeq& - \eta_0 \left[R_{\perp} \left( \frac{d R_{\perp}}{d t}\right)^2 + 
\frac{R_{\parallel}^2}{R_{\perp}}\left( \frac{d R_{\parallel}}{d t}\right)^2 + 
R_{\parallel} \left( \frac{d R_{\perp}}{d t}\right)\left( 
\frac{d R_{\parallel}}{d t}\right) \right] \quad,
\label{Dissip_energy}
\end{eqnarray}
where we have taken into account that typically the velocity gradients are $ \nabla_z  u_x = \nabla_z  u_y \simeq (d R_{\parallel}/d t)/R_{\perp}$  and $ \nabla_x u_z = \nabla_y  u_z \simeq  (d R_{\perp}/d t)/R_{\parallel}$, and the coil volume is $R^{2}_{\parallel}R_{\perp}$.
The relevant friction forces follow from eq. (\ref{Dissip_energy}) by differentiation with respect to the corresponding velocities \cite{Landau1}
\begin{eqnarray}
f_{\rm fric}^{\perp} &=& - \frac{\partial \dot{E}}{\partial\dot{R}_{\perp}}
\simeq \eta_0 \left[ R_{\perp}\frac{d R_{\perp}}{d t} + R_{\parallel} 
\frac{d R_{\parallel}}{d t}\right]
\label{Friction_force}
\end{eqnarray}
\begin{eqnarray}
f_{\rm fric}^{\parallel} &=& - \frac{\partial \dot{E}}{\partial\dot{R}_{\parallel}} 
\simeq \eta_0 \left[ \frac{R_{\parallel}^2}{R_{\perp}} \frac{d R_{\parallel}}{d t} +  
R_{\parallel} \frac{d R_{\perp}}{d t}\right] \simeq f_{\rm fric}^{\perp} \: 
\frac{R_{\parallel}}{R_{\perp}} \: C \quad
%\label{Friction_force}
\end{eqnarray}
where $C$ is a constant whose value will be found later from the initial conditions. 
Note that the ratio of the friction forces is proportional to the aspect ratio of the flattening coil. Since the friction forces are now proportional to one another, in contrast to the case of Rouse dynamics, one of the equations of motion becomes redundant. Thus, after equating the friction and driving forces one derives a single relationship between the perpendicular and the parallel components of the gyration radius, i.e.
\begin{eqnarray}
R_{\parallel}^2(t)= \frac{\sqrt{v N}}{\sqrt{\frac{R_{\perp}}{b^2 N^2} + 
\left( \frac{b^{1/\nu}}{R_{\perp}^{1/\nu + 1}} - \frac{1}{M R_{\perp}}\right) C }}. \quad
\label{Link}
\end{eqnarray}

As usual it is more convenient to use the rescaled variables $x$ and $y$, defined above. In terms of these variables, eq.(\ref{Link}) becomes
\begin{eqnarray}
y^{-1} = \left[ x + \frac{N^2 C}{M^{1 + 2\nu} x}\left( \frac{1}{x^{1/\nu}} - 1\right) 
\right] 
\label{Link_newvar}
\end{eqnarray}
The constant $ C $  in eq.(\ref{Link}) can be fixed by the initial conditions, i.e. $ y = y_0 $ provided that $ x = x_0 $, so that $C = x_0 \left( x_0 - 1/y_0\right) M^{1 + 2\nu}/[\left( 1 - 1/x_0^{1/\nu}\right) N^2]$, and eq. (\ref{Link}) takes the form
\begin{eqnarray}
y^{-1} = x - \frac{x_0 \left( x_0 - y_0^{-1}\right) \left( 1 - x^{-1/\nu}\right)  
}{x \left( 1 - x_0^{-1/\nu}\right) }
\label{Link2}
\end{eqnarray}

As a consequence, only one equation of motion is required for $R_{\perp}$, while $R_{\parallel}$ can then
be obtained directly from  eq. (\ref{Link}). The equation of motion for $R_{\perp}$ can be derived, if one 
notes that the friction force (\ref{Friction_force}) is balanced by the sum of the forces corresponding to 
the free energy contributions corresponding to the attraction $f_{attr}=-\chi_{\infty}N/R_\perp$, and to 
the confinement, $f_{conf}=Nb^{1/\nu}/R_{\perp}^{1/\nu+1}$. One may exclude $dR_{\parallel}/dt$ from eq. 
(\ref{Friction_force}) by taking into account the solvent incompressibility \cite{Landau}, namely 
$\nabla\cdot {\bf u}(r) = 0$, with $ \nabla_x  u_x = \nabla_y  u_y \simeq (d R_{\parallel}/d t)/R_{\parallel}$, 
and $ \nabla_z  u_z \simeq -  (d R_{\perp}/d t)/R_{\perp} $. This yields a simple relation between the two 
velocities and the coil aspect ratio
\begin{eqnarray}
\frac{\dot R_{\parallel}}{\dot R_{\perp}} \simeq \frac{R_{\parallel}}{R_{\perp}}
\label{Aspect ratio}
\end{eqnarray}
Using this relationship in eq. (\ref{Friction_force}), one obtains the equation of motion of $R_\perp$ as:
\begin{eqnarray}
\frac{\zeta_0}{b} \left( R_{\perp} + \frac{R_{\parallel}^2}{R_{\perp}}\right) 
\frac{d R_{\perp}}{d t} = - \frac{1}{M R_{\perp}} + \frac{b^{1/\nu}}
{R_{\perp}^{1/\nu +1}} \quad.
\label{Eq_motion_zimm}
\end{eqnarray}
Making use again of the reduced variables $x$ and $y$, eq. (\ref{Eq_motion_zimm}) can be recast into the form
\begin{eqnarray}
\zeta_0 b^2 \left[ M^\nu x + \left( \frac{v}{b^3}\right)^{1/2} \left( \frac{N}{M^{\nu}}
\right)^{3/2} \frac{y^{1/2}}{x} \right] \frac{d x}{d t} =  \frac{1}{M^{1 + 2\nu} x}
\left( \frac{1}{x^{1/\nu}} - 1\right) \quad.
\label{Eq_motion_zimm22}
\end{eqnarray}

\begin{figure}[ht]
\begin{center}
\epsfig{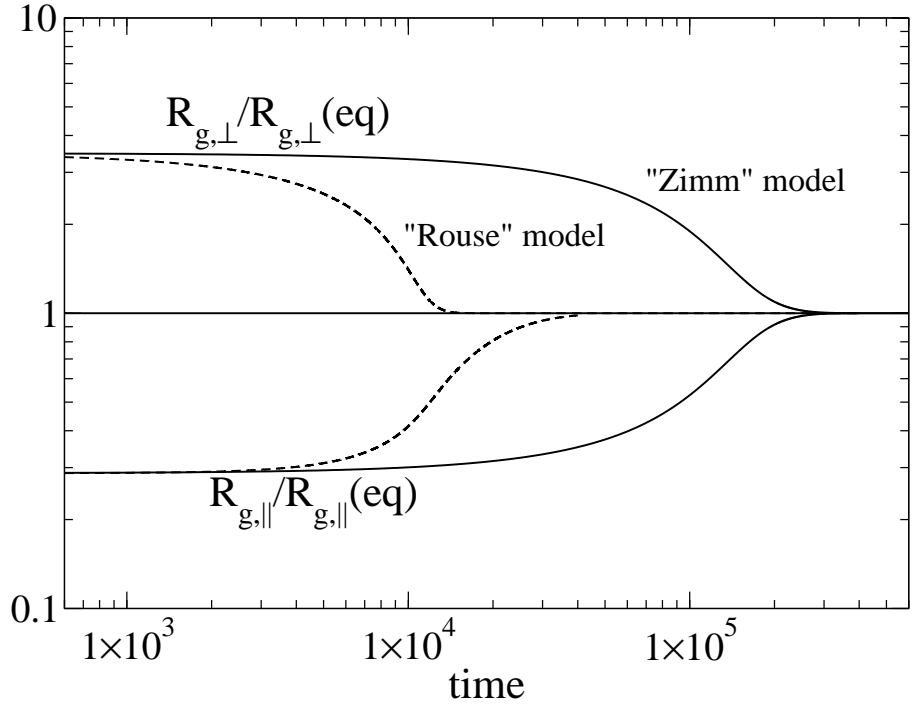}
\vskip 1.0cm
\caption{\label{Rouse_Zimm} 
Log-log plot of the normalized perpendicular and parallel components of the gyration
radius vs time elapsed after the onset of localization for a regular block copolymer 
of length $N=128$ and block size $M=16$ at a flat liquid-liquid interface for the cases 
of Rouse and Zimm dynamics. Here $v=b^3$ and time is dimensionless, measured in units of
$\zeta b^2$.}
\end{center}
\end{figure}

The resulting equation of motion (\ref{Eq_motion_zimm22}), however, in contrast to the case of Rouse dynamics, provides no clear-cut scaling relationship for the typical relaxation time $\tau_{Zimm}$ with respect to $N$ and $M$. Making use of the expression for $y$ in eq.~(\ref{Link2}) and taking the starting conformation as that of an unperturbed self-avoiding coil i.e. $R_{\perp}(0) = R_{\parallel}(0) = b N^{\nu}$, [that is, $ x_0 = \left( N/M\right)^{\nu}$ and $ y_0 = b^3 N^{4\nu} M^{\nu}/(v N^3)$], eq.~(\ref{Eq_motion_zimm22}) can easily be solved numerically for arbitrary chain length $N$ and copolymer block size $M$. 
In Fig. \ref{Rouse_Zimm} results are shown for $N = 128$, $M = 32$, and compared to the predicted localization kinetics for the case of Rouse dynamics, eqs.~(\ref{Eq_motion_x}) and (\ref{Eq_motion_y}). 

Two major conclusions are suggested by a close inspection of Fig.~\ref{Rouse_Zimm}. Evidently, the presence of hydrodynamic interactions in the case of Zimm dynamics tends to erase the  asymmetry in the localization rates perpendicular and parallel to the interface. This is due to the strong coupling between perpendicular and horizontal degrees of freedom induced by the solvent.
In addition, compared to the case of Rouse dynamics, the localization in the presence of hydrodynamic effects takes considerably longer, at least one order of magnitude, under otherwise identical conditions. The physical reason for this slowing down is the presence of high velocity gradients which develop in the incompressible solvent during the coil flattening.

\section{The Simulation Model}

For the Monte Carlo simulations we use an off-lattice bead-spring model that has been employed previously for simulations of polymers both in the bulk and near confining surfaces \cite{KBAM}. Recently, it was applied by us to the study of the static properties of block copolymers at a selective interface \cite{Corsi}, therefore we describe here the salient features only.

Each polymer chain contains $N$ effective monomers connected by anharmonic springs described by the finitely extendible nonlinear elastic (FENE) potential,
\begin{equation} \label{eq1}
U_{FENE} =-\frac{K}{2} R^2 \ln \big[1-\frac{(\ell - \ell_0)^2}{R^2} \big].
\end{equation}
Here $\ell$ is the length of an effective bond such that $\ell_{min}< \ell <\ell_{max}$, with $\ell_{min}=0.4$, $\ell_{max}=1$ being the unit of length, and has the equilibrium value $\ell_0=0.7$, while $R=\ell_{max}-\ell_0=\ell_0-\ell_{min}=0.3$, and the spring constant $K$ is taken as $K/k_BT=40$. The nonbonded interactions between the effective monomers are described by the Morse potential,
\begin{equation} \label{eq2}
U_M=\epsilon_M \{\exp[-2 \alpha(r-r_{min})]-2 \exp [-\alpha(r-r_{min})]\} \,
\end{equation}
where $r$ is the distance between the beads, and the parameters are chosen as $r_{min}=0.8$, $\epsilon_M=1$, and $\alpha=24$. Owing to the large value of the latter constant, $U_M(r)$ decays to zero very rapidly for $r>r_{min}$, and is completely negligible for distances larger than unity. This choice of parameters is useful from a computational point of view, since it allows us to use a very efficient link-cell algorithm \cite{KBAM}.  From a physical point of view, the interactions $U_{FENE}$ and $U_M$, eqs. (\ref{eq1}) and (\ref{eq2}) make sense when one interprets the effective bond as a Kuhn segment, comprising a number of chemical monomers along the chain, and thus the length unit $\ell_{max}=1$ corresponds physically rather to 1 nm than to the length of a covalent $C-C$ bond (which would only be about $1.5 \text{\r{A}}$). Since in the present study we are concerned with the localization of a copolymer in {\em good solvent} conditions, in $U_M(r)$ we retain the repulsive branch of the Morse potential only by setting $U_M(r) = 0 \quad \mbox{for} \quad r > r_{min}$ and shifting $U_M(r)$ upwards by $\epsilon_M$.

The interface potential is taken simply as a step function with an amplitude $\chi$,
\begin{equation}\label{iface}
U_{int}(n,z)=\begin{cases}
-\sigma(n)\chi/2,&z > 0\cr 
\sigma(n)\chi/2,&z \le 0\cr 
\end{cases}
\end{equation}
where the interface plane is fixed at $z = 0$, and $\sigma(n) = \pm 1$ denotes a "spin" variable which distinguishes between P- and H- monomers. In studying the flattening dynamics of these chains, we always start with a configuration which has been equilibrated in a solvent that is good for both types of monomers. At time $t=0$, the interface is switched on (so that it goes through the center of mass of the chain) and the flattening dynamics is then followed until the chain reaches its new equilibrium configuration at the interface. As explained above,  this represents a good initial condition that makes the simulation of the flattening process quite efficient. Since we are interested in the behavior of these chains in the limit of strong localization, we have chosen $\chi=2\chi_{c}$ where $\chi_{c}$ is the crossover selectivity as obtained in Ref. \cite{Corsi}. We use periodic boundary conditions in the plane of the interface while there are rigid walls in the z - direction, where the simulation box extends from $z = -32$ to $z = 32$. Typically we studied chains with lengths $32 \le N \le 512$ and block lengths $1\le M \le N/8$ whereby all measurements have been averaged over $1024$ different and independent equilibrated starting configurations.

\section{Simulation results}

\subsection{The characteristic times of localization}

The results of the Monte Carlo simulations have been compared to the predictions of the Rouse  model introduced above. It is not possible to compare the simulation results to the predictions  of the Zimm model since these simulations do not take into account the hydrodynamic contributions that are so important in the Zimm treatment.

\begin{figure}
\noindent
\begin{minipage}[t]{.49\linewidth}
\epsfig{file=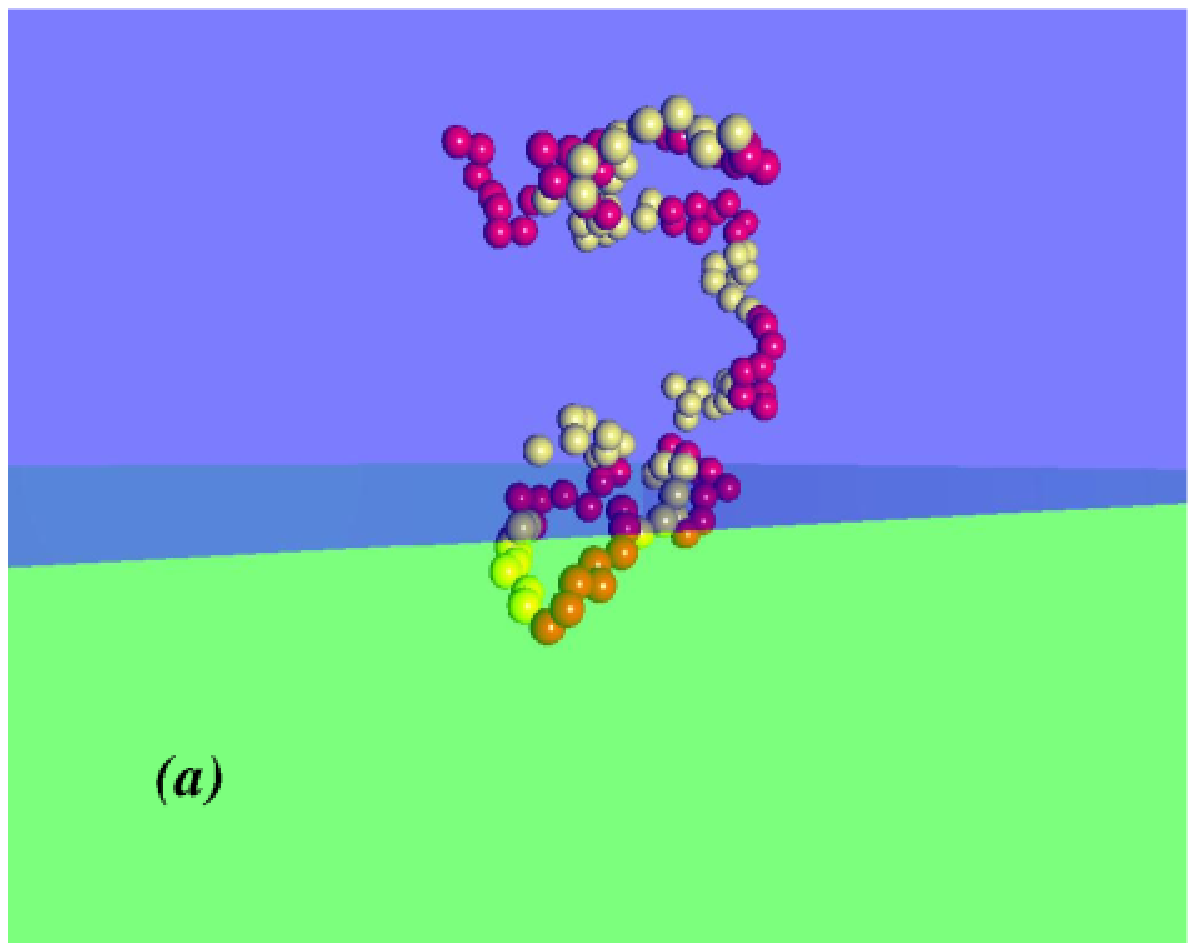,width=1.0\linewidth}
\end{minipage}\hfill
\begin{minipage}[b]{.49\linewidth}
\epsfig{file=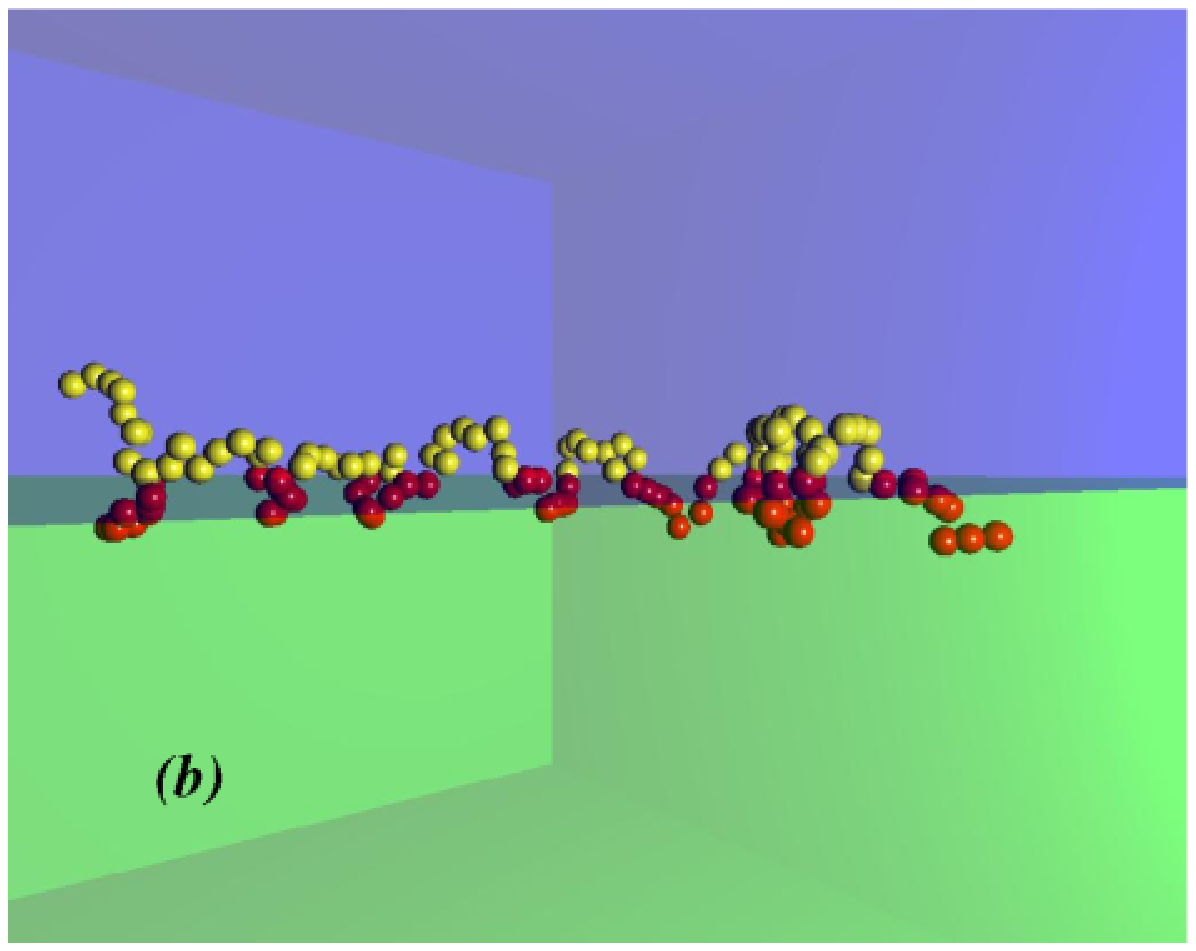,width=1.0\linewidth}
\end{minipage}
%\vskip 1.0cm
\caption{Snapshots of typical configurations of a copolymer with $N = 128$, $M = 8$ at
$\chi = 0.25$ (a), and $\chi = 10$ (b).
The value of the critical selectivity for this chain is $\chi_c = 0.67$. Here the light 
beads prefer the solvent that occupies the upper half of the simulation box, while the 
darker beads prefer the lower half of the box.}
\label{Snapshot}
\end{figure}

Fig. \ref{Snapshot} shows two typical snapshots for a chain with $ N = 128 $ and $ M = 8 $ on 
the verge of the localization threshold ($ \chi = 0.25, \eta = 3.48 $, where $\eta=\chi M^{(1+\nu)/2}N^{(1-\nu)/2}$ is the ``natural'' scaling variable of the problem \cite{Corsi}) and in the strong 
($ \chi = 10, \eta = 139.3 $) localization limit. The value of the crossover selectivity 
for this chain is $\chi_c = 0.67$ \cite{Corsi}.

\begin{figure}[ht]
\begin{center}
\includegraphics[width=11cm]{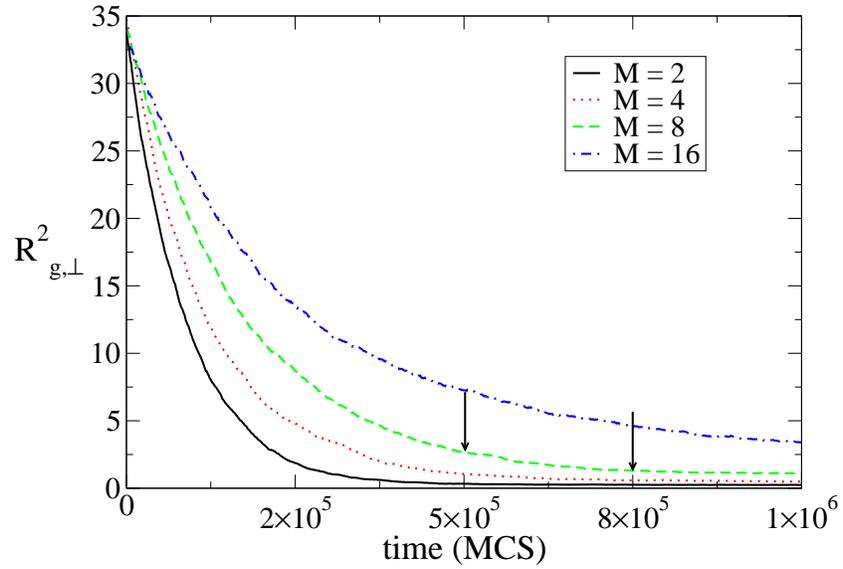}
\caption{\label{R_g}
Typical dependence of $R_\perp (t)$ on time 
in the case of $N = 256$, and $M = 2$ (solid line), $4$ (dotted line), $8$ (dashed
line) and $16$ (dot-dash line).
Arrows denote the time interval where $\tau_\perp$ for the late stage analysis has 
been determined by regression in the case of $M = 8$, see text for details.}
\end{center}
\end{figure}

\begin{figure}[ht]
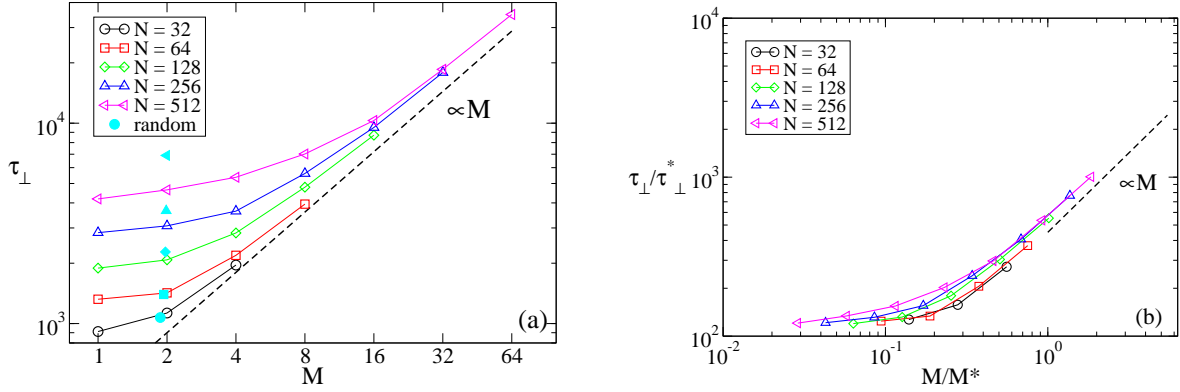

\begin{minipage}[t]{.47\linewidth}
\epsfig{file=initialstages.eps,width=1.0\linewidth}
\end{minipage}\hfill
\begin{minipage}[b]{.47\linewidth}
\epsfig{file=initialstage_master.eps,width=1.0\linewidth}
\end{minipage}
\caption{\label{tau_early} (a)
Variation of the characteristic time for the initial relaxation of $R_\perp$ with
block length $M$ for chains of length $32 \le N \le 512$. The prediction of 
the scaling theory
is shown by a dashed line. The filled symbols represent results for {\em random} 
copolymers with the same values of $N$.
(b) Master scaling plot for the localization of chains of different length $N$ and
block size $M$.
}
\end{figure}

Figure \ref{R_g} shows the typical dependence of $R^{2}_\perp (t)$ on time in the case of $N = 256$, and various $M$ values. These results have been used to determine the dependence of the flattening dynamics on the block length $M$. 
In order to determine the {\it early stages} dynamics, the initial slope of these curves has been obtained and compared with the theoretical prediction of eq. (\ref{Solution_limit11}). 
The results are presented in Fig. \ref{tau_early}. 

From Fig. \ref{tau_early}a one may readily verify that the initial collapse of $R_\perp (t\ll \tau_\perp )$ with 
time closely follows the predicted rate $\propto M^{-1}$ according to eq. (\ref{Solution_limit11}) in the limit 
of $N, M\gg 1$.
Moreover, for sufficiently large block size $M$ the $N$-dependence of this initial rate disappears so that 
asymptotically this initial perpendicular collapse is governed by the length $M$ only, as predicted by the theory.
The disagreement between the prediction of the scaling theory and the results of the simulations  for small 
values of polymer length and block length is not surprising since the predictions  of the scaling theory 
are valid in the limit of $N\to \infty$ and $M\to \infty$, with $N/M$ large. So, we can expect the results of 
the simulations to have a better agreement with the predictions of the theory in the case of very long 
polymers with many long blocks.

We can go one step further in the analysis of the dynamics of the perpendicular component of the radius of 
gyration and assume that the scaling law for the early stages collapse rate has the form
\begin{eqnarray}
\tau_{\perp} = \tau_{\perp}^{*} {\cal G}\left( \frac{M}{M^*}\right) \quad,
\label{Scaling1}
\end{eqnarray}
where $\tau_{\perp}^{*} \propto N^{\alpha_1}  $ and $ M^* \propto N^{\gamma_1} $ and the scaling function
$ {\cal G}(x) $ has the form
\begin{equation}
{\cal G}(x) \left\{\begin{array}{r@{\quad,\quad}l}
1 &x \ll 1\\ x &x \gg 1
\end{array}\right.
\label{Function1}
\end{equation}
According to our prediction, for $ M > M^* $ the $ N $ dependence is dropped out, so that we come to the 
conclusion that $ \alpha_1 = \gamma_1 $. It is possible to obtain the $ N $ - 
dependence of $\tau_{\perp}$ at small $ M $ from Fig. \ref{tau_early}a, and find $ \alpha_1 = 0.57$. When this is known, $ \gamma_1 $ is 
fixed and it is possible to replot Fig. \ref{tau_early}a in terms of $ \tau_{\perp}/\tau_{\perp}^* $ 
and $ M/M^* $ as a master curve - cf. Fig. \ref{tau_early}b. Certainly, this master plot is not
perfect as far as both $N$ and $M$ are still too small for the asymptotic relation, eq. (\ref{Scaling1}),
to be well satisfied but it appears that for growing $N$ the different curves group on a single
one, revealing a clear cut crossover region at $M/M^* \approx 0.3$.

From the curves shown in Fig. \ref{R_g} it is also possible to obtain the characteristic time for the perpendicular component of the radius of gyration corresponding to the late stages of adsorption. To this end the fits of $\tau_\perp$ are taken only {\em after} the initially unperturbed coil has sufficiently relaxed, $R_\perp (t) /R_\perp (0) \le 1/e$. The two vertical lines shown in Fig. \ref{R_g}, for example, indicate the time interval used to fit the late stages behavior for the polymer with total length of $N=256$ and block length $M=8$.

\begin{figure}[ht]
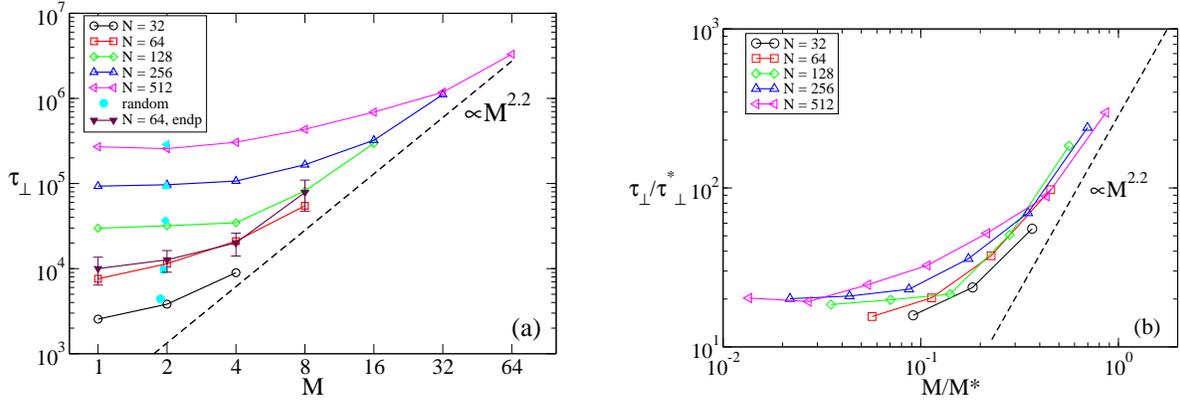
 
\begin{minipage}[t]{.47\linewidth}
\epsfig{file=finalstages.eps,width=1.0\linewidth}
\end{minipage}\hfill
\begin{minipage}[b]{.47\linewidth}
\epsfig{file=finalstages_master.eps,width=1.0\linewidth}
\end{minipage}
\caption{\label{tau_late}
(a) Variation of $\tau_\perp$ with block length $M$ for chains of length $32 \le N \le 512$. The slope of the dashed
line is $\approx 2.2$, according to  eq. (\ref{Solution_limit2}). The filled symbols 
represent results pertaining to {\em random} copolymers with the same chain length $N$. The triangle-down symbols 
represent results corresponding to initial configurations in which the chain barely touches the interface at 
$t=0$. When not explicitly shown, the error bars are smaller than the size of the symbols. (b) Master scaling 
plot for the localization of chains of different length $N$ and
block size $M$.}
\end{figure}

As shown in Fig. \ref{tau_late}a, during the late stages of localization at the interface one observes the expected, 
cf. eq. (\ref{Solution_limit2}), characteristic scaling of $\tau_\perp \propto M^{1+2\nu}$ which progressively 
improves as the asymptotic limit is approached.  As expected, the initial dependence of  $\tau_\perp$ on $N$ 
is seen to vanish as $M$ and $N$ become sufficiently large, in agreement with eq.(\ref{Solution_limit2}).

In Fig. \ref{tau_late}a we also demonstrate (see triangle-down symbols) that the use of initial polymer
configurations which barely touch the interface (rather than having their center-of-mass at $t=0$ at the
interface) does not qualitatively alter the simulational results while it greatly increases the statistical
disarray of the sampling. As mentioned in the beginning of Section 2, this motivated us to choose
in all our computer experiments coils whose center-of-mass is initially placed at the liquid-liquid
interface before the selectivity of both solvents is switched on.

The scaling approach introduced above can also be used for the results concerning the late stages of the localization process 
(Fig. \ref{tau_late}a). In this case the scaling law reads:
\begin{eqnarray}
\tau_{\perp} = \tau_{\perp}^{*} {\cal F}\left( \frac{M}{M^*}\right) \quad,
\label{Scaling2}
\end{eqnarray}
where $\tau_{\perp}^{*} \propto N^{\alpha_2}  $ and $ M^* \propto N^{\gamma_2} $ and the scaling function
$ {\cal F}(x) $ has the following limits
\begin{equation}
{\cal F}(x) \left\{\begin{array}{r@{\quad,\quad}l}
1 &x \ll 1\\ x^{2.2} &x \gg 1
\end{array}\right.
\label{Function2}
\end{equation}
Again the condition that at $ M > M^*$ the $ N $- dependence of $\tau_{\perp}$ is dropped  out leads to the 
link between $ \gamma_2 $ and $ \alpha_2 $; in this case: $ \gamma_2  = \alpha_2/(1+2\nu) = \alpha_2/2.2 $. We can obtain $\alpha_2 = 1.52$ from Fig. \ref{tau_late}a, then determine the value of $\gamma_2$, and finally plot a master curve now in coordinates $ \tau_{\perp}/\tau_{\perp}^* $ versus $ M/M^* $. Fig. \ref{tau_late}b shows the master curve as obtained following this procedure.
Eventually, in Fig. \ref{tau_parall} we display the measured scaling of the late stage characteristic time for the spreading of the chain in the plane of the interface, $\tau_\parallel$, with block size $M$ for different chain lengths $32\le N\le 512$. Apart from some scatter of data for too small ${\cal N} = N/M$, one nicely recovers the relationship $\tau_\parallel \propto N^2$ whereby, once again, the initial $M$-dependence gradually diminishes as ${\cal N} \gg 1$.

\subsection{Rouse mode analysis}

An interesting question about the relationship between localization rate and copolymer composition concerns the mechanism which drives the initially unperturbed coil into a localized state and makes it ``feel'' the presence of a selective interface. A useful insight in this respect may be derived by the detailed Rouse mode analysis of the mode relaxation times.

The Rouse modes are defined as the Fourier cosine transforms of the position vectors, $\bf r_n$, of the repeat unit $n$ of a polymer chain. For the model of discrete polymer chain considered here they can be written in terms of collective coordinates $\bf X_p$ as \cite{Verdier}:
\begin{eqnarray}\label{mode_def}
{\bf X}_p(t) = \frac 1 N \sum_{n=1}^{N}{\bf r}_n(t)\cos\left[
\frac{(n-1/2)p\pi}{N}\right], p=0,1,2,...,N-1
\end{eqnarray} 
The factor $1/2$ is related to the nature of the chain that has free ends. The time-correlation function of the Rouse modes is given by 
\begin{eqnarray}\label{mode_corr}
\Phi_{pq}(t)=\langle{\bf X}_p(t){\bf X}_q(0)\rangle 
\end{eqnarray}
where the angular brackets denote taking an average over different moments of time $t'$ separated by an interval $t$.
For phantom Gaussian chains with no excluded volume interactions one gets
\begin{eqnarray}
\Phi_{pq}^{gauss}(t)= \langle{\bf X}_p(0)
{\bf X}_q(0)\rangle\exp\left(-\frac{t}{\tau_p}\right),\quad p=1,2,...,N-1
\end{eqnarray}
with $(p,q\neq 0)$ coefficients $\langle{\bf X}_p(0){\bf X}_q(0)\rangle_{gauss} 
= \delta_{pq}b^2/[8N\sin^2(p\pi/2N)]$ and relaxation times
\begin{eqnarray}\label{tau_p}
\tau_p^{gauss}=\frac{\zeta b^2}{12k_BT\sin^2(p\pi/2N)}
\end{eqnarray}
so that the typical decay times for the first modes scale as $\tau_p^{gauss}
\propto N^2$. The mode corresponding to $p=0$ describes the self diffusion of the chain. For polymers with excluded 
volume interactions no such analytic results exist but one can show\cite{Doi} that $\tau_p\propto N^{1+2\nu}$. 

\begin{figure}
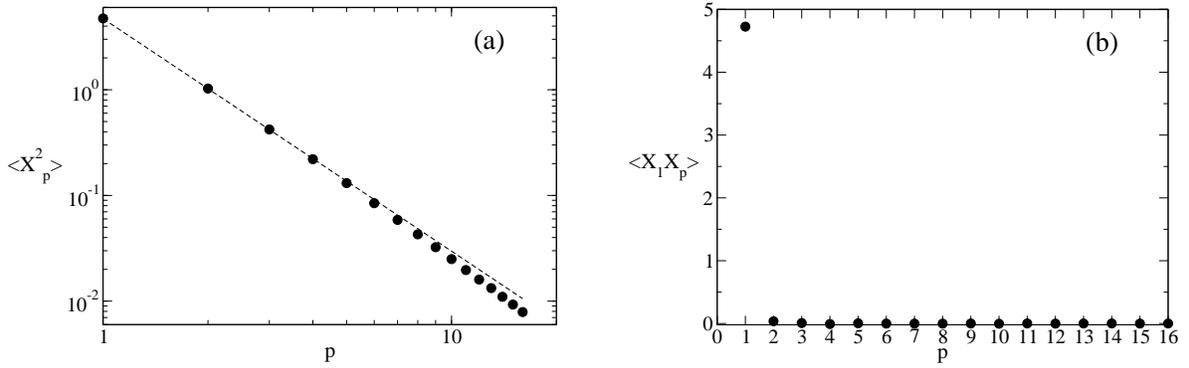

\noindent
\begin{minipage}[t]{.47\linewidth}
\epsfig{file=x_pp_bulk.eps,width=1.0\linewidth}
\end{minipage}\hfill
\begin{minipage}[b]{.47\linewidth}
\epsfig{file=x_1p_bulk.eps,width=1.0\linewidth}
\end{minipage}
\caption{Rouse modes for a chain in the bulk: (a) $\langle{\bf X}_p(0){\bf X}_p(0)\rangle$ 
and (b) $\langle{\bf X}_p(0){\bf X}_1(0)\rangle$ as a function of the mode number $p$ for 
a chain with $N=128$ and $M=16$ in the bulk of a solvent that is good for both kinds of 
monomers. The dashed line in (a) shows the predicted scaling law $\propto p^{-2.2}$.}
\label{Rouse_bulk}
\vspace{1cm}
\end{figure}

It is interesting to examine how the presence of an interface between the two solvents affects the leading modes dynamics. 
In the absence of an interface we find that in equilibrium the first Rouse modes ($1 \le p \le 10$) of a homopolymer chain in a good solvent decay according to expectations, $\tau_p \propto (1/p)^{1+2\nu}\propto N^{1+2\nu}$ and $\langle{\bf X}_p(0){\bf X}_q(0)\rangle \propto (1/p)^{1+2\nu}$. This is shown in Fig. \ref{Rouse_bulk} where the results pertaining to one of the components of the modes are shown for a polymer of length $N=128$ and block length $M=16$. Due to the isotropy of the system, the behavior of the other two components, corresponding to the $y$ and $z$ directions is identical. In part (a) of the figure one can verify that the amplitude of the Rouse modes decays with the mode number as predicted. The discrepancy that occurs for large values of $p$ is expected and it is due to the inability of a coarse-grained model, as the one used in this study, to capture properly the dynamics of the system on too short wavelengths, since the largest $p$ values correspond t!
o motions that occur on very short distances. Fig. \ref{Rouse_bulk}(b) demonstrates the orthogonality of the modes since the only $\langle{\bf X}_p(0){\bf X}_1(0)\rangle$ that is non-zero is the one corresponding to $p=1$. The results shown in 
the figure refer to the $x$ component of the modes, but identical results are obtained if the other two components are plotted. In contrast, for copolymers localized at the liquid-liquid interface one finds a clear asymmetry in their behavior, strongly dependent on the direction that is considered. 

\begin{figure}
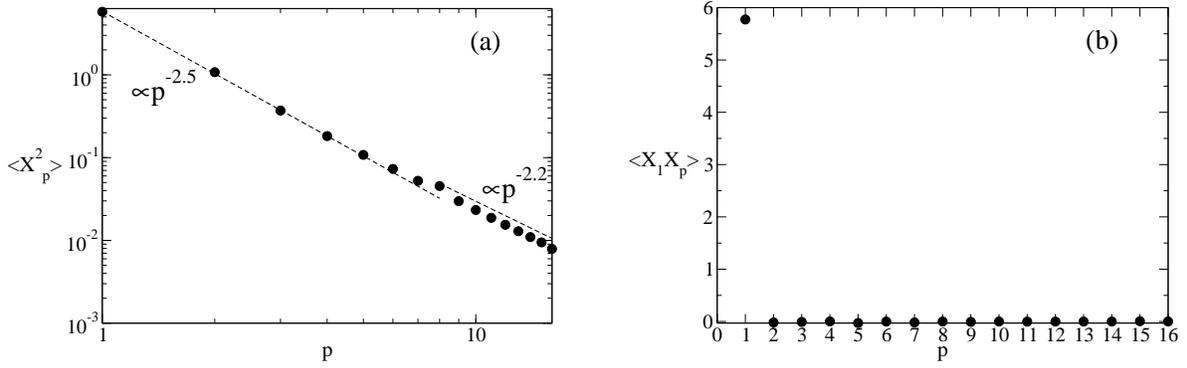

\noindent
\begin{minipage}[t]{.47\linewidth}
\epsfig{file=x_pp_inter.eps,width=1.0\linewidth}
\end{minipage}\hfill
\begin{minipage}[b]{.47\linewidth}
\epsfig{file=x_1p_inter.eps,width=1.0\linewidth}
\end{minipage}
\caption{Rouse modes for a chain at the interface: (a) $\langle X_p(0) X_p(0)\rangle$ and 
(b) $\langle X_p(0) X_1(0)\rangle$ as a function of the mode number $p$ for the components 
of the Rouse modes in the direction parallel to the interface for a chain with $N=128$ and 
$M=16$ strongly adsorbed at a selective interface. The dashed lines in (a) show the predicted 
scaling laws.}
\label{Rouse_inter_parall}
\vspace{1cm}
\end{figure}

Consider first the modes components that are {\em in plane} with the interface, $x$ and $y$. The same quantities, that were plotted in Fig. \ref{Rouse_bulk} for a homopolymer in the bulk, are displayed in Fig. \ref{Rouse_inter_parall} for a copolymer adsorbed at the interface. Again one can see that the decay of the amplitude of the Rouse modes with mode number $p$ still obeys a power-law. However, the polymers are not anymore 3D but rather 2D self-avoiding random walks. Evidently, the relevant Flory exponent in two dimensions $\nu_{2}=3/4$ describes well the dependence of the mode amplitudes on the mode numbers as demonstrated by the dashed line with slope $\propto -(1+2\nu_{2})= -2.5$ in Fig. \ref{Rouse_inter_parall}. It is very interesting to notice how the modes corresponding to large values of p follow the power-law decay observed in the three-dimensional case. The modes with $p<8$ describe monomer motions which involve more than $M$ particles. Thus the modes corresponding to long wavelengths which span more than one block length feel the presence of the interface in contrast to those corresponding to shorter wavelengths (and show a dependence like $\propto p^{-(1+2\nu)}$, see Fig. \ref{Rouse_inter_parall}).
In part (b) of the same figure, we show that the Rouse modes components parallel to the interface are still orthogonal to each other. The only non-zero $\langle{X}_p(0){X}_1(0)\rangle$  is the one corresponding to $p=1$. This holds for both the $x$ and $y$ components of the modes. Careful inspection of the figure reveals that the value of $\langle{X}_1(0){X}_1(0)\rangle$ is almost twice larger than that of the bulk case shown in Fig. \ref{Rouse_bulk}. This is a consequence of the increased in-plane fluctuations that the polymer chain exhibits when it is confined to a plane.

\begin{figure}
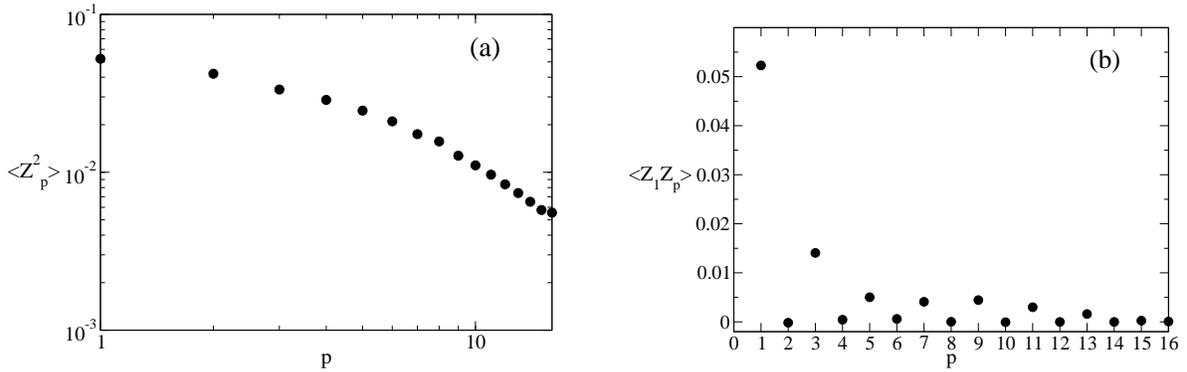

\noindent
\begin{minipage}[t]{.47\linewidth}
\epsfig{file=z_pp_inter.eps,width=1.0\linewidth}
\end{minipage}\hfill
\begin{minipage}[b]{.47\linewidth}
\epsfig{file=z_1p_inter.eps,width=1.0\linewidth}
\end{minipage}
\caption{Rouse modes for a chain at the interface: (a) $\langle Z_p(0) Z_p(0)\rangle$ and 
(b) $\langle Z_p(0) Z_1(0)\rangle$ as a function of the mode number $p$ for the components 
of the Rouse modes in the direction perpendicular to the interface for a chain with $N=128$ 
and $M=16$ strongly adsorbed at a selective interface.}
\label{Rouse_inter_perp}
\vspace{1cm}
\end{figure}

If we look at the component of the Rouse modes corresponding to the direction perpendicular to the interface, the results are quite different as shown in Fig. \ref{Rouse_inter_perp}. In Fig. \ref{Rouse_inter_perp}(a) one finds no power-law dependence of $\langle{Z}_p(0){Z}_p(0)\rangle$ on the mode number $p$. The amplitudes of the perpendicular components of the modes decrease weakly with $p$ and are several orders of magnitudes smaller than those of the parallel components of the modes. Evidently, the strong interface supresses all fluctuations of the chain perpendicular to the interface which is manifested by the strong decrease in the amplitude of the perpendicular components of the modes.

Another striking difference in comparison to the results observed for the parallel components, can be seen in Fig. \ref{Rouse_inter_perp}b. The modes are {\em not orthogonal} to each other anymore since the quantities $\langle{Z}_p(0){Z}_1(0)\rangle$,  corresponding to $p=3,5,7$ up to $p=15$, are non-zero and of the same order of magnitude as $\langle{Z}_1(0){Z}_1(0)\rangle$. Thus one can conclude that the odd-numbered modes are coupled to one another. The amplitude of these couplings is so small, though, that the system does not lose its ergodicity. The same can be observed for the even-numbered modes too but the amplitude of these terms is even smaller than the amplitude observed for the odd-numbered modes.

\begin{figure}
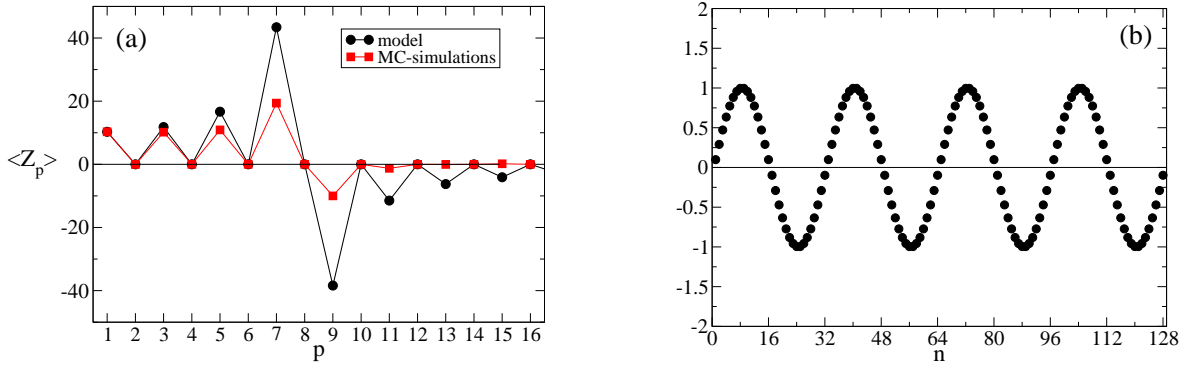

\noindent
\begin{minipage}[t]{.46\linewidth}
\epsfig{file=mathematical_model_comparison.eps,width=1.0\linewidth}
\end{minipage}\hfill
\begin{minipage}[b]{.43\linewidth}
\epsfig{file=mathematical_model.eps,width=1.0\linewidth}
\end{minipage}
\caption{(a) Comparison between the average values of the perpendicular components of the 
Rouse modes obtained from the trial function, see eq. (\ref{model}), and the values obtained 
from Monte Carlo simulations for a chain with $N=128$ and $M=16$ strongly adsorbed at a 
selective interface; (b) the trial function used to describe the average conformation of 
a chain with $N=128$ and $M=16$.}
\label{Surface_model}
\vspace{1cm}
\end{figure}

Another observation concerns the {\em average values} of the perpendicular components of the Rouse modes. If one calculates $\langle{X}_p\rangle$ or $\langle{Y}_p\rangle$, then due to the isotropy of the system in the plane of the interface these values vanish for all $p$'s because there is no preferential orientation in the plane. If one calculates $\langle{Z}_p\rangle$, instead, the results are not trivial. The values obtained for a polymer of length $N=128$ and block length $M=16$ are shown in Fig. \ref{Surface_model}a. The squares represent the results obtained from the Monte-Carlo simulations of such chains. A pattern is clearly present in the $\langle{Z}_p\rangle$: all the values corresponding to {\em even} $p$'s vanish 
while those corresponding to odd $p$'s are non zero. The values of $\langle{Z}_p\rangle$ for odd $p$ increase with growing $p$, and they change sign for $p>N/M$. In order to test if we were able to describe this behavior simply by the localization pattern at the interface, we approximated the shape of the localized chain perpendicular to the interface as a simple sine wave,
\begin{eqnarray}\label{model}
Z(n)=A(M)\sin\left[\frac{(n-1/2)\pi}{M}\right],
\end{eqnarray} 
where $n$ is an index that numbers the monomers in the chain ($n=1..N$), and $A$ is a prefactor related to the size of the adsorbed polymer in the direction perpendicular to the interface. This simplest sinusoidal shape, see Fig. \ref{Surface_model}b, is periodic with the block length $M$ while the first and last monomers of the chain are not pinned at $z=0$ accounting for the loose first and last blocks with respect to the interface. 
Using eq. (\ref{model}) in eq. (\ref{mode_def}) to obtain $\langle{Z}_p\rangle$, we get what is represented by the circles in Fig. \ref{Surface_model}a. Note that we have rescaled the Monte Carlo results so that the two sets of data start from the same value for $p=1$. Evidently, our simple expression for the {\it average} shape of the chain is able to reproduce remarkably well the simulation results, and in particular, to catch the trend in the amplitude growth as well as the change in sign observed at $p=N/M$.
One may therefore conclude that the observed behavior of the Rouse modes for the chains adsorbed at the interface is indeed a consequence of the interplay between the interface and the regular block structure of the polymers under investigation.  

\begin{figure}
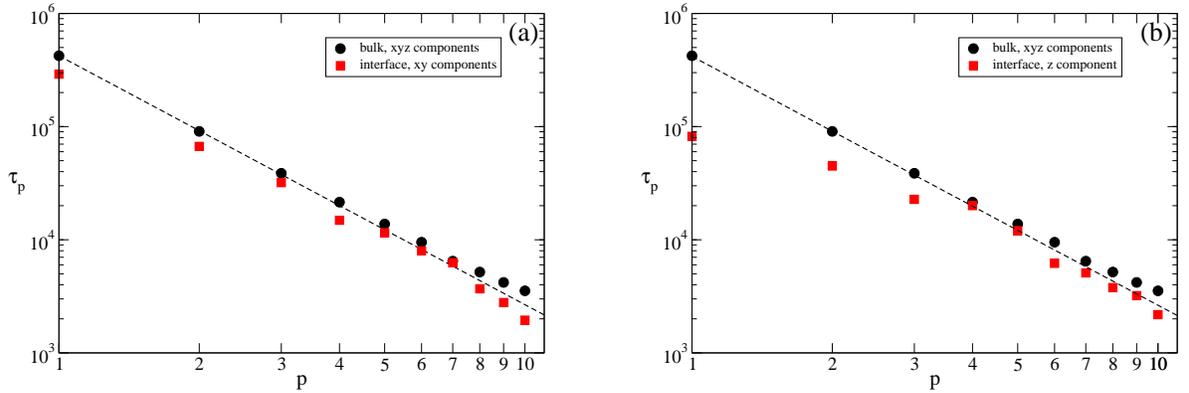

\noindent
\begin{minipage}[t]{.46\linewidth}
\epsfig{file=tau_xy.eps,width=1.0\linewidth}
\end{minipage}\hfill
\begin{minipage}[b]{.46\linewidth}
\epsfig{file=tau_z.eps,width=1.0\linewidth}
\end{minipage}
\caption{Relaxation times of the Rouse modes: (a) comparison between the bulk behaviour and the behaviour of the components parallel to the interface for the localized chains; (b) comparison between the bulk behaviour and the behaviour of the component perpendicular to the interface for the localized chains. As above, the results shown were obtained using a chain with $N=128$ and $M=16$ strongly adsorbed at a selective interface.}
\label{Rouse_decay}
\vspace{1cm}
\end{figure}

Eventually, we end our discussion of the Rouse-mode behavior with an analysis of the Rouse time spectrum - cf. Fig. \ref{Rouse_decay}. 
In both parts of Fig. \ref{Rouse_decay} we plot the relaxation times of the first ten Rouse modes for a polymer chain in the bulk (black circles in the figure). The dashed line represents the expected theoretical dependence of $\tau_{p}$ on $p$ as $\sim p^{-2.2}$. As it can be easily observed, the results of the simulations follow the theoretical prediction quite nicely. As in the case of the amplitude of the Rouse modes, one observes deviations from the predicted law when the condition $p/N \ll 1$ does not hold anymore.
We compare the results for the chain in the bulk to the respective results for chains localized at the interface. In Fig. \ref{Rouse_decay}(a), we consider the components parallel to the interface. Apparently the presence of the interface does not affect the relaxation time $\tau_p$ considerably. Evidently, the times $\tau_p$ for the chain at the interface are shorter than those for the chain in the bulk so that the presence of the interface effectively speeds up the inplane dynamics. 
%Also, it seems that the mode corresponding to $p=4$ is the one that has changed the most relative to the bulk value. We think this might be related to the fact that the mode with $p=4$ is commensurate to the block length and we would expect a similar effect for $p=8,16...$. Unfortunately the error in the results grows dramatically as $p$ increases and it becomes impossible to check this for sure. 
In Fig. \ref{Rouse_decay}(b), instead, the bulk behavior is compared to the behavior of the perpendicular component at the interface. In this case the times $\tau_p$ corresponding to small values of $p$ change more significantly. The presence of the interface accelerates the relaxation of the modes that are characterized by large wavelengths (small values of $p$) since then the chain can  better ``sense'' the interface. 
%On the contrary, the effect on the modes with larger $p$ seems much less remarkable, as expected also considering what we have discussed above. Once again, the erratic behavibehaviore results for very large values of $p$ are just related to a not good enough statistics and not to intrinsic reasons.

\subsection{Localization of random copolymers}

Having understood the impact of block structure on copolymer localization kinetics, it is tempting to contrast it to the case of {\em random} HP-copolymers and the rate of their flattening onto a liquid-liquid interface. In recent years there has been a number of studies of random copolymers both by analytical theory \cite{Sommer1,Garel,Joanny,Denesyuk} as well as by means of computer simulation \cite{Chen,Peng,Chen2}. As a rule, all these studies focus on the static conformational properties of random copolymers at the penetrable interfaces between two solvents. Thus it is now well established \cite{Sommer1,Peng} that the perpendicular component of the gyration radius $R_\bot (\chi)$ of a random copolymer should scale with the degree of interface selectivity as a function of $\eta=\chi\sqrt{N}$ for small and moderately large values of the interface selectivity $\chi$.

\begin{figure}[ht]
\begin{center}
\includegraphics[width=11cm] {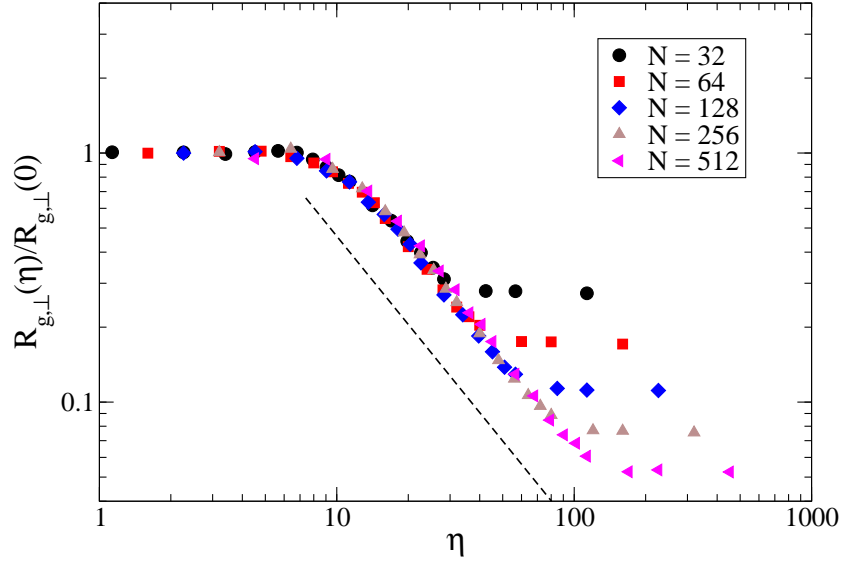}
\caption{\label{random} $R_\bot(\eta)/R_\bot(0)$ as a function of the scaling variable 
$\eta = \chi\sqrt{N}$ for different chain lengths. The dashed line represents the predicted 
scaling behavior in the weak localization regime, $\propto \eta^{-2\nu}$.} 
\end{center}
\end{figure}

\begin{figure}[ht]
\begin{center}
\includegraphics[width=11cm] {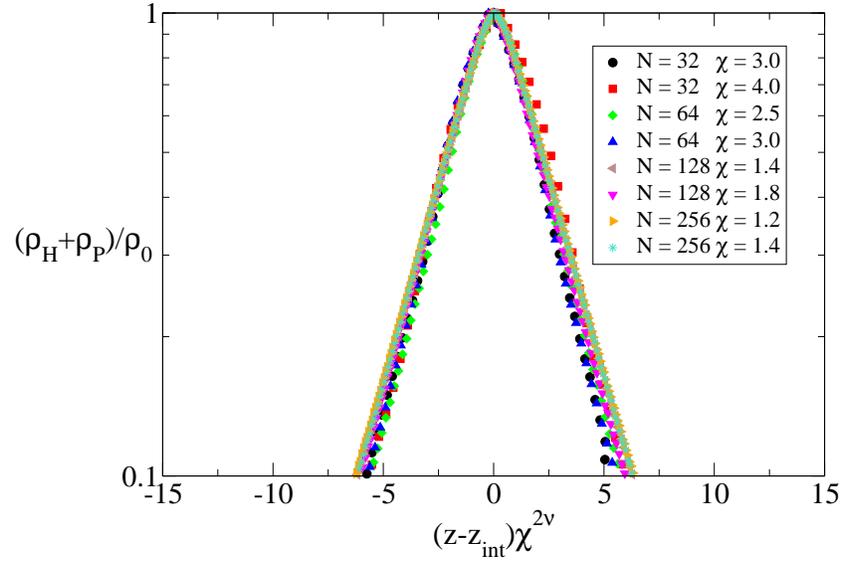}
\vskip 0.9cm
\caption{\label{rand_dens} Scaling plot of the total monomer density in terms of 
$(z-z_{int})\chi^{2\nu}$ for different combinations of polymer length and interface 
strength.} 
\end{center}
\end{figure}

In order to determine the crossover selectivity $\chi_{cr}$ in our model, we carried out simulations on random copolymers with length $32\le N\le 512$ and the results may be seen in the scaling plot of Fig. \ref{random} where all data are found to collapse onto a single curve. One may estimate from the master curve in Fig. \ref{random} that $\chi_{cr}\sqrt{N}\approx 20$. The dashed line in Fig. \ref{random} shows the predicted \cite{Sommer1, Peng} dependence of the perpendicular component of the radius of gyration on the scaling variable $\propto \eta^{-2\nu}$ in the weak localization regime. As extensively explained elsewhere \cite{Corsi}, the scaling is expected to fail when $\chi$ is very large because then the blob size reaches its minimum possible value corresponding to a single block. Mathematically, the scaling fails for large values of the selectivity because of the different normalization factors used for the different curves as $R_\bot(0)$ depends of course on $N$. We have also checked another scaling hypothesis put forward for the first time in \cite{Peng}: the total polymer density in the region of the interface should scale with $\chi^{2\nu}$ when $\chi$ lies within the limits of the weak localization regime: 
\begin{eqnarray}\label{rand_dens_scaling}
\rho_{P}(z)+\rho_{H}(z)=\rho_{0}g[(z-z_{int})\chi^{2\nu}]
\end{eqnarray}
where $\rho_{P}(z)$ is the density of $P$ monomers at a distance $z-z_{int}$ from the interface, where $z_{int}$ is the position of the interface ($z_{int}=0$ in our simulations), $\rho_{H}(z)$ is the corresponding quantity for the $H$ monomers, $\rho_{0}$ is the total monomer density at the interface, and $g(y)$ is a function close to being exponential.
Fig. \ref{rand_dens} demonstrates that this scaling law is indeed observed for a wide range of polymer lengths and interface strength values.

\begin{figure}[ht]
\begin{center}
\epsfig{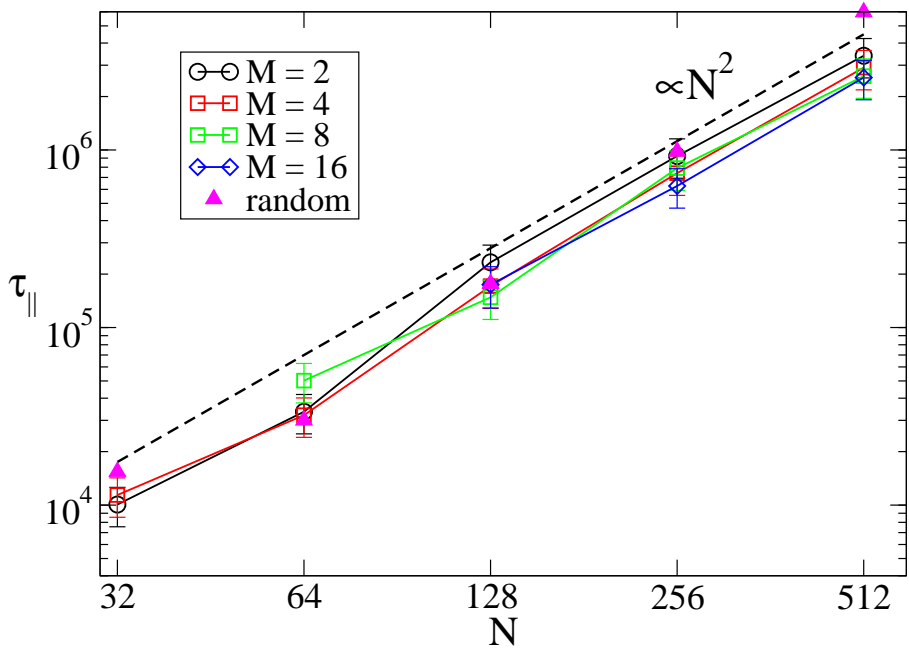}
\caption{\label{tau_parall}
Variation of $\tau_\parallel$ with chain length $N$ for blocks of size $M = 2, 4, 8$ and $16$. 
The slope of the dashed line is 2, as predicted by the theory. The filled triangles represent 
results for random copolymers with the same values of $N$.}
\end{center}
\end{figure}

Also in the case of random copolymers, we performed all our studies of the localization kinetics in the limit of strong localization. It can be easily proved that, in the limit of infinite chain length, the average block size in a random copolymer is $\langle M \rangle = 2$. In the case of polymers with finite length, there are finite-size corrections that have to be taken into account \cite{Corsi2}: the average block size is then always smaller than $2$, albeit very close to this value, so it is appropriate to compare the rate of flattening of random copolymers to that of a multiblock copolymer with block size $M=2$ at the same selectivity strength $\chi = 2\chi_{cr}$. 
The results for the flattening dynamics of random copolymers are compared to those for regular block copolymers in Figures \ref{tau_early} to \ref{tau_parall} where the results corresponding to the random copolymers are represented by filled symbols. In these figures it is possible to notice a marked difference between the early stage and the late stage dynamics. As seen in Fig. \ref{tau_early}a, the values of $\tau_\perp$ for the random copolymers and the corresponding block copolymers are very close for small values of $N$. As $N$ increases, however, the difference between these two values grows as well and is particularly large in the limiting case of $N=512$ when the characteristic time for the random copolymer is about twice as large as the value for the corresponding block copolymer. A careful inspection of Fig. \ref{tau_early}a shows that the value of $\tau_\perp$ for a random copolymer with $N=512$ is similar to that of a regular block copolymer of the {\em same} total length and a considconsiderablyr block size $M=8$. 

The trend shown by the simulation results as the length of the polymers increases can be easily understood through a careful analysis of the nature of the random sequences that can be created in a polymer of finite length \cite{Corsi2}: when the polymer length is short, e.g. $N=32$, the probability of generating sequences of consecutive like monomers much longer than $2$ is extremely small. As a matter of fact, the polymer will almost always contain a sequence of $H-$ and $P-$ monomers in which the length of each segment is small since every single polymer used to run the simulation is prepared in a {\em neutral} state with as many $P-$ as $H-$monomers. On the contrary, when a very long polymer is generated, the probability of having longer sequences of consecutive $H-$ or $P-$ monomers is increased. A long random copolymer will likely have long sequences of like monomers at some point along the chain. These long sequences are responsible for the much faster adsorption kinetics as compared to that of the regular block copolymers.
The presence of long sequences of like monomers along the chain apparently affects only the early stages of the localization dynamics: from Figures \ref{tau_late}a and \ref{tau_parall}a it can be observed that during the late stages of the adsorption process, the behavior of random copolymers is almost identical to that of the corresponding regular block copolymers with $M=2$. This observtiobservationunderstood by considering that during the late stages of adsorption the long blocks that are present in the random copolymers have already been adsorbed at the interface and it is the remaining short blocks that drive the system in order to complete the adsorption process. During this stage it is then the shortest blocks that are important for the adsorption process, and in this respect the random copolymers are very similar to the regular blocky ones, with blocks of length equal to $2$.

\section{Conclusions}

From the results presented above, it appears that the simple scaling theory that we have developed to describe the kinetics of copolymer adsorption on a selective liquid-liquid interface is able to capture quite well the most salient features of the problem as a comparison with extensive MC simulation data demonstrates. The agreement between the predictions of the model and the results of the simulations in both limits of early and late stages of adsorption is remarkable especially for long polymer and block lengths. 

The analysis that we have carried out regarding the Rouse modes of the system indicates that polymer chains that are strongly absorbed at the interface behave as two dimensional random walks in the plane of the interface, and show almost no fluctuations in the direction perpendicular to the interface. A simple trial function for the shape of the chain in the direction perpendicular to the interface is sufficient to capture the features of the modes that are specific to the adsorbed chains.

Finally, we have also shown how the random copolymers behave at the interface. In particular, we have been able to prove how the presence of long blocks in the random sequences of the polymer chains affecs maffectsthe early stages of adsorption while the behavior of the random chains in the late stages of adsorption is almost identical to that of regular block copolymers with $M=2$, as expected.

Although the present study deals with the simplest situation of symmetric interface selectivity and copolymer
composition and it assumes equal viscosity of both immiscible liquids, it
can be viewed as a first step in exploring the localization kinetics of 
copolymers at the liquid-liquid interface. We also believe that our results could stimulate exciting 
laboratory experiments and be tested, for example, by means of the method of videomicroscopy
of single macromolecules\cite{Gurrieri}.
\vskip 1cm

\underline{\em Acknowledgments} AM acknowledges the support and hospitality of the Max-Planck Institute for Polymer Research in Mainz during this study. This research has been supported by the Sonderforschungsbereich (SFB 625).

\newpage

\newpage

\newpage

\newpage

\vspace {4.0cm}

\newpage
 
\vspace {4.0cm}

\newpage
 
\newpage
 
\newpage 
 
\newpage
 
\newpage
 
\newpage
 
\newpage
 
\newpage

\newpage
 
\vspace{3cm}
\newpage
 
\begin{thebibliography}{99}
\bibitem{Clifton} B. J. Clifton, T. Cosgrove, R. M. Richardson, A. Zarbakhsh, and
J. R. P. Webster, Physica B {\bf 248}, 289 (1998)
\bibitem{Rother}G. Rother, and G. F. Findenegg, Colloid Polym. Sci. {\bf 276}, 496 (1998).
\bibitem{Wang} R. Wang, and J. B. Schlenoff, Macromolecules {\bf 31}, 494 (1998).
\bibitem{Omarjee} P. Omarjee, P. Hoerner, G. Riess, V. Cabuil, and O. Mondain-Monval,
Eur. Phys. J. E {\bf 4}, 45 (2001).
\bibitem{Sommer1} J.-U. Sommer and M. Daoud, Europhys. Lett. {\bf 32}, 407 (1995).
\bibitem{Sommer2} J.-U. Sommer, G. Peng and A. Blumen, J. Phys. II (France) {\bf 6},
1061 (1996).
\bibitem{Garel} T. Garel, D.A. Huse, S. Leibler and H. Orland, Europhys. Lett. {\bf 8},
9 (1998).
\bibitem{Joanny} X. Chatellier and J.-F. Joanny, Eur. Phys. J. E {\bf 1}, 9 (2000).
\bibitem{Denesyuk} N.A. Denesyuk and I.Ya. Erukhimovich,  J. Chem. Phys. {\bf 113}, 
3894 (2000).
\bibitem{Balasz}A. C. Balasz, and C. P. Semasko, J. Chem. Phys. {\bf 94}, 1653 (1990).
\bibitem{Israels} R. Israels, D. Jasnow,  A.C. Balazs , L. Guo, G. Krausch, J. Sokolov, and
M. Rafailovich, J. Chem. Phys. {\bf 102}, 8149 (1995).
\bibitem{Sommer3}J.-U. Sommer, G. Peng and A. Blumen, J. Chem. Phys. {\bf 105}, 8376 (1996).
\bibitem{Lyats} Y. Lyatskaya, D. Gersappe, N.A. Gross, and A.C. Balazs,
J. Chem. Phys. {\bf 100}, 1449 (1996).
\bibitem{Chen} Z. Y. Chen, J. Chem. Phys. {\bf 111}, 5603 (1999);
{\bf 112}, 8665 (2000).
\bibitem{Corsi}
A. Corsi, A. Milchev, V.G. Rostiashvili, and T.A. Vilgis, J. Chem. Phys. {\bf 122}, 
094907 (2005).
\bibitem{Grossberg} A.~Yu.~Grossberg and A.~R.~Khokhlov, in {\em Statistical
Physics of Macromolecules} AIP Press, N. Y., 1994. 
\bibitem{Landau} L.~D.~Landau and E.~M.~Lifshitz, in {\em Course of Theoretical
Physics}, Vol. 6, Fluid Mechanics, Butterworth-Heinemann, 3rd edition, Oxford, 1987.
\bibitem{Landau1} L.~D.~Landau and E.~M.~Lifshitz, in {\em Course of Theoretical
Physics}, Vol. 1, Mechanics, Butterworth-Heinemann, 3rd edition, Oxford, 1987.
\bibitem{Leclerc} E. Leclerc and M. Daoud, Macromolecules {\bf 30}, 293 (1997).
\bibitem{Gennes} P.-G. de Gennes,  {\em Scaling Concepts in Polymer Physics}, Cornell
University Press, N.Y., 1979.
\bibitem{KBAM} K. Binder and A. Milchev, J. Computer-Aided Mater. Design {\bf 9}, 33 (2002).
\bibitem{Verdier} P.~H.~Verdier, J. Chem. Phys. {\bf 45}, 2118 (1966).
\bibitem{Doi} M.~Doi and S.~F.~Edwards, {\em The Theory of Polymer Dynamics} (Clarendon,
Oxford, 1986).
\bibitem{Peng} G.~W.~Peng, J.~U.~Sommer, and A.~Blumen, Phys. Rev. E {\bf 53}, 5509 (1996). 
\bibitem{Chen2} Z. Y. Chen, J. Chem. Phys. {\bf 112}, 8665 (2000).
\bibitem{Corsi2}
A. Corsi, A. Milchev, V.G. Rostiashvili, and T.A. Vilgis, {\em to be published} 
\bibitem{Gurrieri} S. Gurrieri, B. Smit, S. Wells, D. Johnson, and C. Bustamante,
Nucleic Acids Res., {\bf 24}, 4759 (1996).
\end{thebibliography}
\end{document}